\documentclass[preprint,12pt]{elsarticle}

\usepackage{amssymb}
\usepackage{amsmath}

\usepackage{pythonhighlight}

\newcounter{bla}

\journal{Computer Physics Communications}

\begin{document}

\begin{frontmatter}

\title{DiPolMol-Py: A Python package for calculations for $^{2}{\Sigma}$ ground-state molecules}

\author[a]{Bethan Humphreys}
\author[a]{Alex J. Matthies}
\author[a]{Hannah J. Williams\corref{author}}

\cortext[author] {Corresponding author.\\\textit{E-mail address:} hannah.williams4@durham.ac.uk}
\address[a]{Department of Physics, Durham University, South Road, Durham DH1 3LE, United Kingdom}

\begin{abstract}
We present the python package DiPolMol-Py, which can be used to calculate the rotational and hyperfine structure of $^2\Sigma$ molecules.  The calculations can be performed in the presence of dc magnetic fields, dc electric fields and far off-resonant optical fields. We additionally include functions to calculate the polarisability of the molecule and the transition dipole moment between different energy eigenstates. The package is applicable to many of the molecules which can be laser cooled, specifically the alkaline earth fluorides. We provide a constants file which includes many of the required literature values for CaF, SrF and BaF. Additional species can easily be added by updating this file.

\noindent \textbf{PROGRAM SUMMARY}

\begin{small}
\noindent
{\em Program Title:} DiPolMol-Py                                     \\
{\em CPC Library link to program files:} (to be added by Technical Editor) \\
{\em Developer's repository link:} https://github.com/durham-qlm/DiPolMol \\
{\em Licensing provisions:} BSD 3-clause \\
{\em Programming language:} Python $\geq$3.11                                   \\
{\em Nature of problem:} Calculating the rotational and hyperfine structure for $^2\Sigma$ ground state molecules both field free and in the presence of dc magnetic, electric and off-resonant light fields.\\
{\em Solution method:} A Python package which calculates the eigenenergies and eigenvalues via diagonalisation of the Hamiltonian.\\
{\em Additional comments including restrictions and unusual features (approx. 50-250 words):}\\
  This package is based on previous work for $^1\Sigma$ molecules~[1]. External magnetic and electric fields must be coaxial.
   \\

\end{small}
   \end{abstract}
\end{frontmatter}

Ultracold molecules offer many exciting opportunities from fundamental science to quantum technologies, made possible by their complex internal structure. Transitions between internal energy levels can be used as sensitive probes to search for variations in fundamental constants or beyond standard model physics, e.g. electrons electric dipole moment~\cite{Lim2018,Aggarwal2018,Anderegg2023,Roussy2023}, nuclear Schiff moment~\cite{Grasdijk2021}, and electron-to-proton mass ratio~\cite{Barontini2022}. The long-range, tuneable dipole-dipole interactions present between rotational levels allow for the entanglement of molecules~\cite{Yan2013,Holland2023,Bao2023,Ruttley2024,Picard2024}, which combined with long-coherence times~\cite{Williams2018,Burchesky2021, Park2023, Gregory2024} make molecules a promising platform for quantum computing and simulation~\cite{DeMille2002,Carr2009,Hazzard2013,Hazzard2014,Cornish2024}. The rotational level structure extends the capabilities of such platforms to include synthetic dimensions~\cite{Sundar2018, Feng2022} or qudits~\cite{Sawant2020}. With the ability to control internal and external degrees of freedom, cold molecules are also ideal for the study of cold, controlled chemistry~\cite{Toscano2020,Heazlewood2021,Liu2022} and collisions~\cite{Cheuk2020,Jurgilas2021,Koller2022}. 

Direct laser cooling of molecules is one way of producing ultracold samples of molecules. It was first proposed in 2004~\cite{DiRosa2004}, and since then the field has advanced rapidly. The first molecular magneto-optica trap (MOT) was demonstrated in 2014 using strontium monofluoride (SrF)~\cite{Barry2014}, with several further species being trapped since (CaF~\cite{Truppe2017,Anderegg2017}, YO~\cite{Collopy2018}, CaOH~\cite{Vilas2022}, SrOH~\cite{Lasner2024}, $^{138}$BaF~\cite{Zeng2024}). Even more are being investigated currently (YbF~\cite{Alauze2021}, $^{137}$BaF~\cite{Kogel2021}, CaH~\cite{VazquezCarson2022}, CaD~\cite{Dai2024}, MgF~\cite{Pilgram2024}, AlF~\cite{Hofsass2021}, TlF~\cite{Norrgard2017}, CH~\cite{Schnaubelt2021}).

Implementing any of the potential applications of molecules requires an accurate understanding of the internal energy level structure, both with and without the presence of external fields. A Python package \textit{diatom.py}~\cite{Blackmore2023} was recently written to calculate the rotational and hyperfine structure of $^1\Sigma$ molecules, applicable typically to associated, bialkali molecules. In this paper we present a complementary Python package \textit{DiPolMol.py} which works for $^2\Sigma$ molecules, and is thus applicable to many laser-coolable diatomic molecules. We provide most\footnote{Constants for SrF and BaF light shift and polarisability could not be found in literature.} of the required constants to calculate the structure of CaF, SrF and BaF. As more species are investigated, and the required constants become known, these can be added to the programme. This could include associated molecules such as RbSr~\cite{Ciamei2018}, RbYb~\cite{Franzen2023}, CsYb~\cite{Guttridge2018} or LiYb~\cite{Green2019} which similarly have a $^2\Sigma$ ground state.

The remainder of this paper is structured as follows. Section (Theory) outlines the Hamiltonian used in the calculations; considering rotational and hyperfine structure as well as the effect of external magnetic and electric fields. Section (Package) outlines how the code is structured and how the user can perform calculations. Finally, section (Examples) demonstrates an application of the DiPolMol-Py package to CaF, SrF and BaF to calculate polarisability, Zeeman shifts, ac and dc electric field shifts and transition dipole moments between different rotational states.

\section{\label{theory}Theory}

In this package we consider a molecule in the ground electronic and vibrational state. The total Hamiltonian describing such a molecule is
\begin{equation}
	H_{\rm tot} = H_{\rm rot} + H_{\rm hf} + H_{\rm ext},
	\label{eq:Ham}
\end{equation}
where $H_{\rm rot}$ and $H_{\rm hf}$ give the field-free rotational and hyperfine structure, respectively, and $H_{\rm ext}$ describes the interaction with external electric, magnetic and off-resonant light fields. We give the Hamiltonian in Eqn.~\ref{eq:Ham} using dimensionless operators for electron spin, $\hat{\boldsymbol{S}}$, nuclear spin, $\hat{\boldsymbol{I}}$ and rotational angular momentum $\hat{\boldsymbol{N}}$. In the package we use the matrix representation of the Hamiltonian to calculate the associated energy shifts. The matrix representation for each Hamiltonian is presented in the Appendix.

\subsection{\label{basis}Basis}
The electronic ground state for most molecules which are being pursued as candidates for laser cooling is $^2\Sigma$. The term symbol is given by $^{2S+1}\Lambda$, where $S$ is the total electronic spin angular momentum and $\Lambda$ is the projection of the total electronic orbital angular momentum $L$ onto the internuclear axis. Values of $\Lambda =  0, 1, 2,...$ are represented by $\Lambda =\Sigma, \Pi, \Delta,...$ Thus, molecules in $^2\Sigma$ states have $S=\frac{1}{2}$ and $\Lambda=L=0$. 

\begin{figure}
	\centering
	\includegraphics[width=0.75\linewidth]{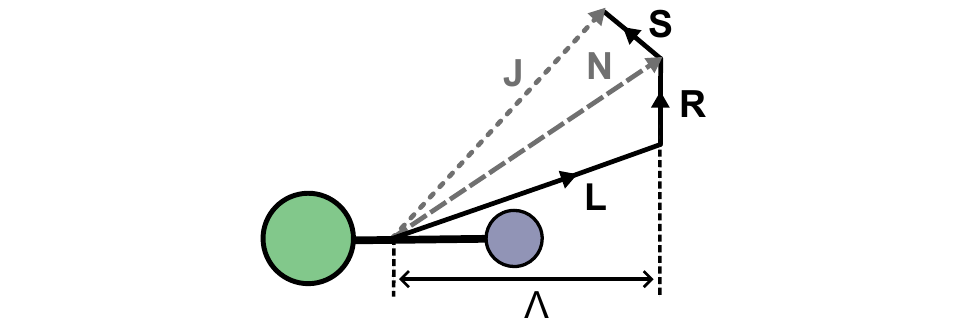}
	\caption{Illustration of Hund's case (b) angular momentum couplings. The orbital angular momentum $\mathbf{L}$ and the rotational angular momentum $\mathbf{R}$ couple to form $\mathbf{N}$, which in turn couples to the spin angular momentum $\mathbf{S}$, forming $\mathbf{J}$. The projection of the orbital angular momentum onto the internuclear axis is given by $\Lambda$, as shown.
    }
	\label{fig:Hunds}
\end{figure}

Hund's case (b) best describes this system whereby the rotational angular momentum $R$ is first coupled to the electronic angular momentum $L$ to make $N =L+R$. $N$ is then coupled with electronic spin $S$ to give the total angular momentum $J = N+S$. The coupling scheme is illustrated in figure \ref{fig:Hunds}. Note that as we consider molecules in the $^2\Sigma$ state, $N$ only has a contribution from the rotational angular momentum, i.e.,  $N = R$. The total angular momentum $J$ couples with the nuclear spin $I$, to give the hyperfine structure with $F = J + I$. The projection of the angular momentum $F$ onto the internuclear axis is given by $m_{F}$. The natural state basis is thus the coupled regime $|N, J, F, m_F \rangle $.

\subsection{Rotational Hamiltonian}
The molecular rotation can be approximated by the rigid-rotor model, resulting in an array of rotational states spaced in energy according to $E_N \approx B_v N(N+1)$, where $B_v$ is the rotational constant in vibrational level $v$. However, this approximation does not hold for large N. At  faster rates of rotation there is an increase in the centrifugal force which causes the bond length to grow. This effect is captured by the centrifugal distortion coefficient $D_v$, which is typically six orders of magnitude smaller than $B_v$. The rotational Hamiltonian is~\cite{BandC}
\begin{equation}
	H_{\rm rot} = B_v(\hat{\boldsymbol{N}}\cdot \hat{\boldsymbol{N}}) - D_v(\hat{\boldsymbol{N}}\cdot \hat{\boldsymbol{N}})^2.
\end{equation}

\subsection{Hyperfine Hamiltonian}
In this package, we consider only those molecules where one of the constituent atoms has no nuclear spin, i.e., for a molecule AB $I_{\rm A}=0$ and $I_{\rm AB} = I_{\rm B}$. This is the case for molecules containing $^{40}$Ca, $^{88}$Sr or $^{138}$Ba and covers the majority of molecules currently being investigated for laser cooling. The hyperfine Hamiltonian takes the form 

\begin{equation}
	H_{\rm hf} = H_{\rm e~spin-rot} + H_{\rm spin-spin} + H_{\rm n~spin-rot}. 
\end{equation}

The first term describes the electronic spin-rotation interaction and is given by
\begin{equation}
    H_{\rm e~spin-rot} = \gamma \hat{\boldsymbol{S}}\cdot \hat{\boldsymbol{N}},
\end{equation}
where $\gamma$ is the electron spin-rotational coupling constant.

The second term describes the interaction between the electronic and nuclear magnetic moments and can be decomposed into a scalar and a tensor part. These are written as 
\begin{equation}
    H_{\rm spin-spin}^{(0)} = (b+c/3)\hat{\boldsymbol{I}}\cdot \hat{\boldsymbol{S}},    
\end{equation}
\begin{equation}
    H_{\rm spin-spin}^{(2)} = (c/3)\sqrt{6} T^2(C)\cdot T^2(\hat{\boldsymbol{I}},\hat{\boldsymbol{S}}),
\end{equation}
with spectroscopic constants $b$ and $c$~\cite{Frosch1952}. $T^2(C)$ and $T^2(\hat{\boldsymbol{I}},\hat{\boldsymbol{S}})$ are rank-2 spherical tensors where $(C)$ represents the renormalised spherical harmonics $C_q^2(\theta,\phi)$.

The final term is the nuclear spin-rotation interaction, written as

\begin{equation}
	H_{\rm n~spin-rot} = c_{\rm F} \hat{\boldsymbol{I}}\cdot \hat{\boldsymbol{N}}.
\end{equation}
The nuclear interaction is typically three orders of magnitude smaller than the others, with constant $c_{\rm F}$~\footnote{Sometimes referred to as $C$.}.

\subsection{External Hamiltonian}
We next consider the interaction between the molecule and an external electromagnetic field. We introduce $\boldsymbol{\lambda}$ as a unit vector along the internuclear axis in the direction from negative to positive charge (e.g., from the F to the Ca in CaF). The total external Hamiltonian can be expressed as the sum of three Hamiltonians which describe the effect of magnetic, electric and off-resonant optical fields. The total external Hamiltonian is therefore

\begin{equation}
	H_{\rm ext} = H_{\rm B} + H_{\rm dc} + H_{\rm ac}.
\end{equation}

An external magnetic field $\boldsymbol{B}$ leads to a splitting in energy levels described by the Zeeman Hamiltonian, which can be expressed as~\cite{BandC}
\begin{eqnarray}
	H_{\rm B} = g_{\rm s}\mu_{\rm B} \hat{\boldsymbol{S}} \cdot \boldsymbol{B} + g_{\rm l}\mu_{\rm B}[\hat{\boldsymbol{S}} \cdot \boldsymbol{B}-(\hat{\boldsymbol{S}} \cdot \boldsymbol{\lambda} )(\boldsymbol{B}\cdot\boldsymbol{\lambda})]\nonumber\\ 
	- g_{\rm r}\mu_{\rm B} \hat{\boldsymbol{N}} \cdot \boldsymbol{B} - g_{\rm N}\mu_{\rm N}\hat{\boldsymbol{I}} \cdot \boldsymbol{B},~~~~~~~~
\end{eqnarray}
where the first term is generally three orders of magnitude larger than the others. The four terms describe the contributions of the electron's magnetic dipole moment, the anisotropic correction to this, the rotation of the electron, and the nuclear magnetic moment respectively. These terms are characterised by their associated $g$-factors $g_{\rm s},g_{\rm l},g_{\rm r}$ and $g_{\rm N}$.\

The permanent electric dipole moment $\mu_{\rm e}$ of a molecule couples to an external dc electric field $\boldsymbol{E}_{\rm dc}$. The Hamiltonian describing this interaction is
\begin{equation}
	H_{\rm dc} = -\mu_{\rm e}\boldsymbol{E}_{\rm dc}\cdot\boldsymbol{\lambda}.
\end{equation}

Finally, we include the interactions between the molecule and a non-resonant light field. Here the dipole moment operator of the molecule $\hat{\boldsymbol{d}}$ interacts with the ac electric field of the light $\boldsymbol{E}_{\rm ac}$.  The associated Hamiltonian is written as
\begin{equation}
    H_{\rm ac} = -\hat{\boldsymbol{d}}\cdot\boldsymbol{E}_{\rm ac}.
\end{equation}

When considering optical fields it is generally simplest to think about the intensity of the light, which is related to the square of the electric field. The intensity $\mathcal{I}$ is given by
\begin{equation}
    \mathcal{I} = \frac{c\epsilon_{\rm{0}}}{2}{\left|\mathbf{E_{ac}}\right|}^2.
\end{equation}

The size of the light shift depends on the polarisation of the light and the frequency-dependent polarisability of the molecule $\alpha$, which consists of a scalar, vector and tensor part $\alpha_k$, $k={0,1,2}$. The scalar polarisability leads to an equal shift of all levels, which can be expressed as 
\begin{equation}
    U_{\rm{0}} = - \frac{\alpha_0}{2 \epsilon_{\rm{0}} c} \mathcal{I} = - \alpha_0' \mathcal{I}. 
\end{equation}
For simplicity we have introduced the reduced polarisability $\alpha_k'=\alpha_k/(2c\epsilon_0)$. 
 
\section{Package}
The package contains two main notebooks. The first \texttt{Hamiltonian}, is used to build the Hamiltonians and the second \texttt{Calculate} uses eigenstates and eigenenergies found from diagonalising the Hamiltonian to run various calculations. We also include a file \texttt{Constants} which includes all of the required constants for CaF~\cite{Kaledin1999,Dagdigian1974,Wall2008}, as well as the constants required to calculate the field-free, Zeeman and dc electric field effects for BaF~\cite{Chen2016} and SrF~\cite{Dagdigian1974,Berg1996,Childs1981}.

\subsection{Hamiltonian}
We construct the Hamiltonian as a 2d array of size $d$,
\begin{equation}
d = (2I + 1)\times (2S+1) \times \sum_{N=0}^{N_{\rm max}}(2N+1), 
\end{equation}
where $N_{\rm max}$ is defined by the user as the maximum rotational level to be included in the calculation. A Hamiltonian is constucted using the \texttt{build} function, which takes \texttt{N\textsubscript{max}} (integer), \texttt{Constants} (dictionary) and \texttt{Zeeman, dc, ac} (Booleans) as parameters. The \texttt{Constants} term requires the importing of the chosen molecular species dictionary, e.g., `CaF' from the constants file. The \texttt{Zeeman, dc, ac} must be set to either `True' or `False' depending on which external fields are needed for the calculation. If an external field value is zero, the time required for a calculation can be reduced by setting the associated Boolean to `False'. The full Hamiltonian can then be constructed by simply adding together $H_0 = H_{\rm rot} + H_{\rm hf}$ with the external Hamiltonians multiplied by their respective fields, i.e., magnetic field with $H_{\rm B}$, electric field with $H_{\rm dc}$ and light intensity with $H_{\rm ac}$. An excerpt of code to calculate the Zeeman energies for SrF is given in the Examples section.

\subsection{Calculate}
The calculate file contains a selection of functions which can be used to identify quantum numbers of eigenstates and calculate transition dipole moments and polarisabilities.

\subsubsection{State identification}

The function \texttt{np.linalg.eigh} calculates the eigenstates and eigenenergies of the Hamiltonian and outputs these in order according to eigenenergy from lowest to highest. If two levels cross in energy, then this can lead to a misidentification of eigenstates. To overcome this we include the function \texttt{sort\_smooth}~\cite{Blackmore2023}, which takes the arrays of eigenenergies and eigenstates calculated over a range of external field values as an input. Eigenstates are tracked by ensuring maximal overlap as the external field is changed, and the eigenenergies are reordered as appropriate to match.

To assign the sorted eigenstates with appropriate $F$ and $m_F$ labels, the function \texttt{label\_FmF\_states} 
is provided within \texttt{calculate}. It takes the sorted set of eigenstates as an input, along with \texttt{N\textsubscript{max}} (integer), \texttt{Constants} (dictionary) and the magnetic field at which the eigenstates and eigenenergies are  calculated, \texttt{B} (integer/float/list/array) \footnote{A non-zero B field must be applied to lift the Zeeman sublevel degeneracies}, and returns the relevant $(N, F, m_F)$  or $(N, F)$ labels. 

We first generate a list of possible $|N,J,F,m_F\rangle$ basis states, following the order these quantum numbers are cycled through when constructing the maxtrix representation of the Hamiltonian in the \texttt{hamiltonian} function. The \textit{$i^{th}$} entry in a given eigenstate can therefore be assigned as the coefficient of the \textit{$i^{th}$} basis state in this list. For each eigenstate, all basis states with negligible coefficients are removed. In the case when multiple basis states remain, these are checked to ensure consistency in $F$ and $m_F$ quantum numbers. If this condition is met, these values are returned as the $F, m_F$ label for that eigenstate. In the case that $m_F$ is not a good quantum number, only $F$ labels will be output.

For simplicity, the function \texttt{solve} is also provided, combining the functionality of \texttt{np.linalg.eigh}, \texttt{sort\_smooth} and optionally, \texttt{label\_FmF\_states}. As inputs, this function requires \texttt{H} (np.array), the matrix representation of the Hamiltonian to solve, alongside \texttt{N\textsubscript{max}} (integer) and \texttt{Constants} (dictionary). $(N, F, m_F)$ labels can also be generated by setting the argument \texttt{label} (Boolean) to \texttt{True} and providing an additional input, \texttt{B} (integer/float/list/array), the magnetic field value(s) at which the eigenstates and eigenenergies are to be calculated. 

\begin{figure}
	\includegraphics[width=0.75\textwidth]{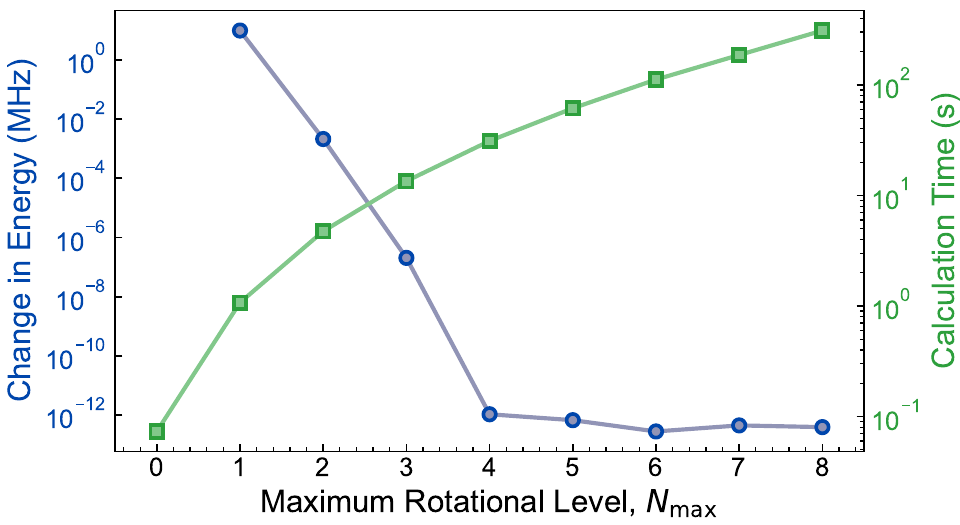}
    \centering
	\caption{Plot showing the effect on the accuracy of energy calculation for $|N=0, F=0, m_F=0\rangle$ (blue circles) and the time to run calculation (green squares) as the maximum rotational level $N_{\rm max}$ considered is increased, up to 8 for CaF. The calculation is done with $\mathcal{I}=2.5\times10^9$~W~m$^{-2}$, $E=50$~kV~m$^{-1}$ and $B=200$~G.} 
	\label{fig:Nmax}
\end{figure}

\subsubsection{Transition dipole moments}
A second function within \texttt{calculate} is \texttt{transition\_dipole\_moment} which calculates the transition dipole moment between two states. The function takes \texttt{N\textsubscript{max}, Constants, M, states} and \texttt{gs} as inputs. Here, \texttt{gs} is the chosen ground state, \texttt{states} are the other states being considered, and \texttt{M} is the helicity of the transition such that $\texttt{M}=+1, 0, -1$ for $\sigma^+, \pi, \sigma^-$ transitions respectively.
The function outputs the transition dipole moment in units of the molecule frame dipole moment $d_0$. The first step is to calculate the induced dipole moment operator $\boldsymbol{\mu}$, using function \texttt{dipole}, the matrix representation of which is the same as for the dc electric field effect. We then calculate the expectation value between a chosen ground state \texttt{gs} and every other state via matrix multiplication using \texttt{np.einsum}.

\subsubsection{Polarisability}
We include a function \texttt{alpha\_012} which calculates the scalar ($\alpha'_{0}$), vector ($\alpha'_{1}$) and tensor ($\alpha'_{2}$) components of the reduced molecular polarisability at a given wavelength.

We make the assumption that only the lowest two electronic states i.e., $A ^{2}\Pi$ and $B ^{2}\Sigma^+$ contribute to the polarisability and don't include higher lying states in the calculation~\cite{Caldwell2020}. The function takes the wavelength of the light and \texttt{Constants} as inputs and generates a list of $\alpha_k'$ for $k=0,1,2$ following the equations given in the Appendix.

\subsection{Number of rotational states considered}

In order to accurately determine eigenenergies, it is important to include an appropriate number of rotational levels, especially when considering large electric fields which cause rotational mixing. However, increasing the number of levels, $N_{\rm max}$, included in the calculation also increases the size of the associated Hilbert space and hence the length of time required for a calculation. The dependence of the calculated eigenenergies on $N_{\rm max}$ is shown in figure \ref{fig:Nmax}. The blue circles show the change in the energy of the lowest energy state $|N=0, F=0, m_F=0\rangle$ as  $N_{\rm max}$ is increased $\Delta E = (E(N_{\rm max})-E(N_{\rm max}-1))/h$. This is calculated for $\mathcal{I}=2.5\times10^9$~W~m$^{-2}$, $E=50$~kV~m$^{-1}$ and $B=200$~G. Once $N_{\rm max}\ge 4$ the change in energy is less than $10^{-12}$~MHz with the inclusion of an additional rotational level. Also shown in figure \ref{fig:Nmax}, as green squares, is the time taken to run these calculations on a laptop with 11th Gen Intel(R) Core(TM) i5-1135G7 @ $2.40$~GHz with $32$~GB of RAM. To achieve the required accuracy with $N_{\rm max}=4$, the calculation time is still reasonable at around 30 seconds.

\begin{figure*}
	\includegraphics[width=\linewidth]{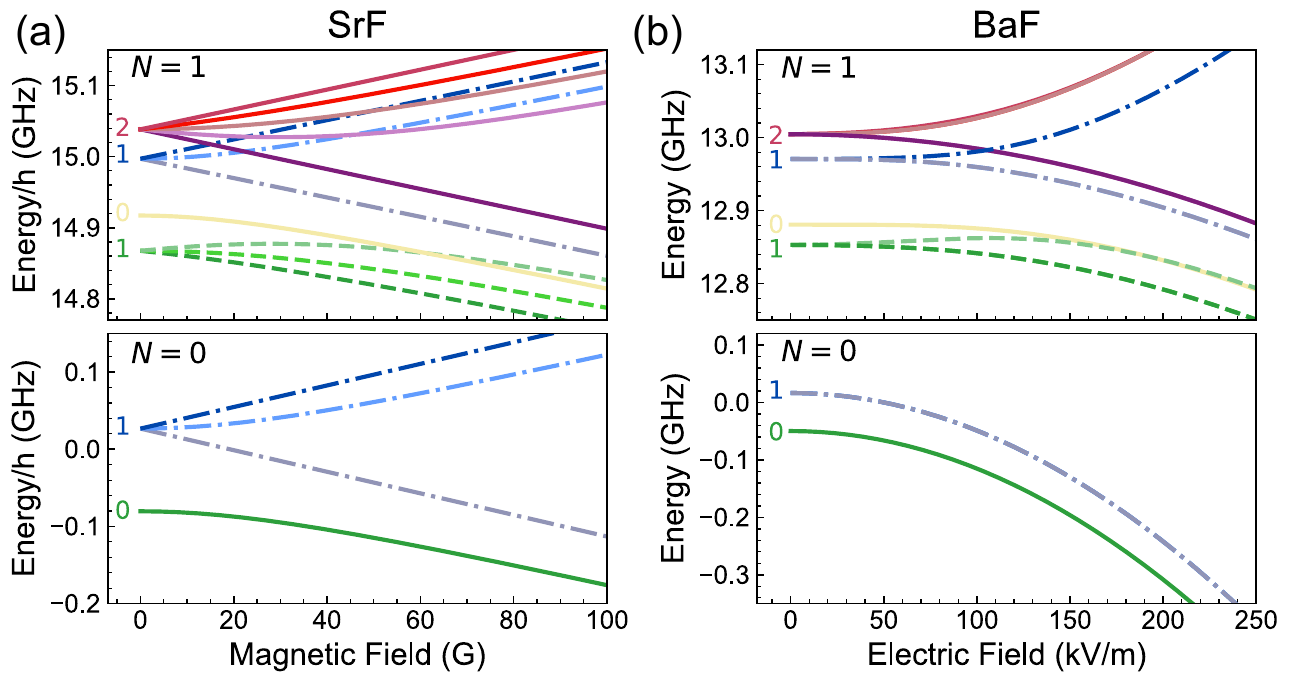}
	\caption{Plots showing (a) Zeeman effect for SrF molecules and (b) dc electric field shift for BaF molecules. The $N=0$ states are shown in the lower panels and $N=1$ states are shown in the upper panels. In each case only the plotted field is applied. The calculations include states up to $N=4$. The line styles are set according to their $F$ quantum number (labelled by the coloured number) and $m_F$ levels are identified by the colour scale.}

	\label{fig:ZeeDC}
\end{figure*}
\section{Examples}
This section includes an excerpt of the code used to build and diagonalise a Hamiltonian and example energy calculations.

\subsection{Zeeman effect and dc electric field shift}
The first example considers the Zeeman effect for SrF. The Hamiltonian is constructed to use only the Zeeman contribution, the magnetic field values are defined and then applied to the Hamiltonian. The eigenstates, eigenenergies and state labels are then calculated using the \texttt{calc.solve} function. The code is as follows:

\begin{python}
	import numpy as np
	import hamiltonian as hamiltonian
	import calculate as calc
	from constants import SrF

	Nmax=4 #Identify the maximum N 
	H0,H_B,H_dc,H_ac 
		= hamiltonian.build
			(Nmax,SrF,zeeman=True,Edc=False
              ,Eac=False) 

	B = np.linspace(0,100,5000)*1e-4 #Tesla
	
	H = H_0[..., None] + H_B[..., None]*B
	H = H.transpose(2,0,1)
	
	energies, states, label_list = 
       calc.solve(H, Nmax, SrF,label=True, B)

\end{python}

This code produces the data shown in figure \ref{fig:ZeeDC}(a). The magnetic field is varied from $0$ to $100$~G in steps of $2$~mG, with no electric or light fields present. The $N=0$ and $N=1$ rotational levels are shown in separate panels as the rotational splitting is $\approx 15$~GHz. In both panels, the splitting of the different $m_F$ states can be clearly seen.

Similarly, the dc electric field shifts can be calculated by slightly adjusting the above code. Namely, changing the arguments passed to \texttt{build} such that \texttt{zeeman=False, Edc=True}, as well as replacing \texttt{H\_B} with \texttt{H\_dc} in the definition of the variable \texttt{H}.
The results of this calculation for BaF are shown in figure \ref{fig:ZeeDC}(b). Here, the electric field is varied from $0$ to $250$~kV~m$^{-1}$ in steps of $150$~V~m$^{-1}$, with no magnetic or light field present. Again, the results are separated into two panels according to rotational number. Comparing these panels to those in figure \ref{fig:ZeeDC}(a), it can be seen that the dc electric field does not lift the degeneracy between all $m_F$ states.

\begin{figure*}
    \centering
    \includegraphics[width=\linewidth]{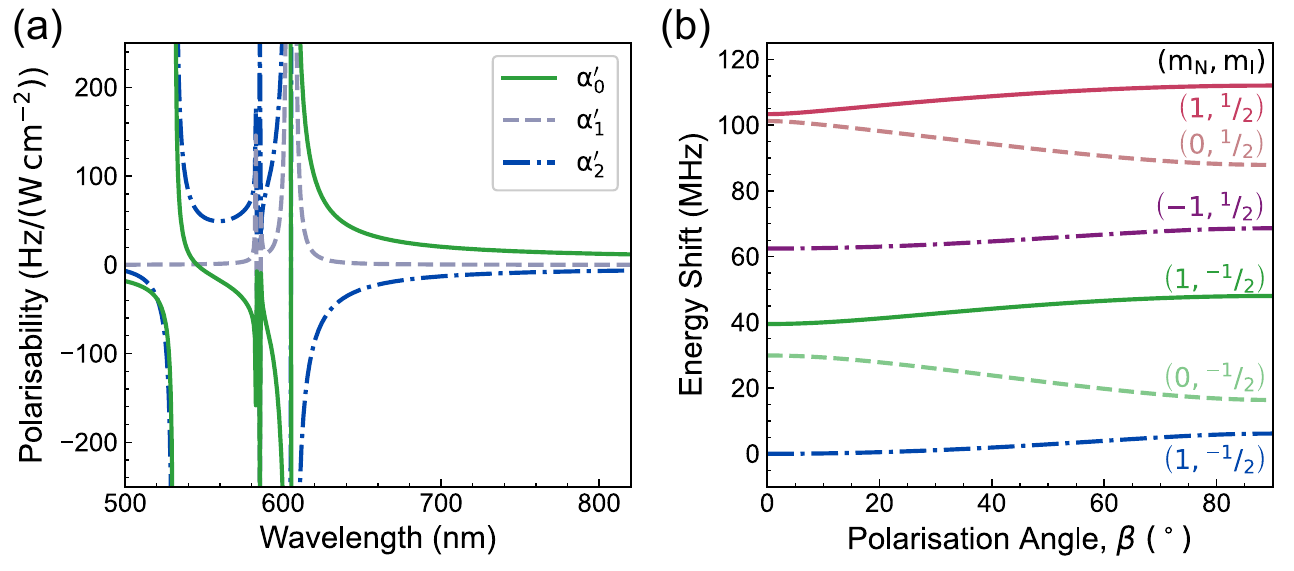}
    \caption{(a) Calculated polarisabilities for CaF in the ground electronic and vibrational state, $X(v=0)$. The scalar component ($\alpha'_{0}$) is shown in the solid green line, the vector component ($\alpha'_{1}$) in the dashed grey line and the tensor component ($\alpha'_{2}$) in the dash-dotted blue line. (b) ac light shift for CaF in the $N=1, m_s=-1/2$ levels. Calculations are performed with a magnetic field of $B=300$~G, and for linearly polarised 780~nm light at an angle $\beta$ to the B field and an intensity of $\mathcal{I}=30$~GW~m$^{-2}$. The states are labeled by the $m_{\rm N}$ and $m_{\rm I}$ quantum numbers.
    Zero energy is defined to be the energy of the lowest energy state at $\beta=0^\circ$}
    \label{fig:pol_ac}
\end{figure*}

\subsection{Polarisability and light shift}
The effect of off-resonant light is becoming increasingly important to understand as molecules are regularly being loaded into conservative optical traps, such as optical dipole traps and tweezers~\cite{Anderegg2019,Holland2023}. To compute the light shift, we need the values of the polarisability components $\alpha_k^{'}$, which can either be set in the \texttt{constants} file or calculated using the \texttt{polarisability} function for a given wavelength. Figure \ref{fig:pol_ac}(a) shows the three polarisability components calculated for CaF, using the \texttt{polarisability} function over a wavelength range of $\lambda = 500$~nm to $\lambda = 820$~nm.

Additionally, the light shift can be calculated similarly to the Zeeman and dc electric field shifts by modifying the \texttt{build} function. Figure \ref{fig:pol_ac}(b) shows an example for CaF molecules. The calculations were done using linearly polarised light field of wavelength $780$~nm, intensity $30$~GW~m$^{-2}$ with a magnetic field of 300~G and no electric field. We plot the energy shift against the angle ($\beta$) between the polarisation of the light and the applied magnetic field from 0 to 90$^\circ$ relative to the lowest energy state at $\beta=0$. Here we identify the levels by $m_N$ and $m_I$. 
\begin{figure}
	\includegraphics[width=0.55\textwidth]{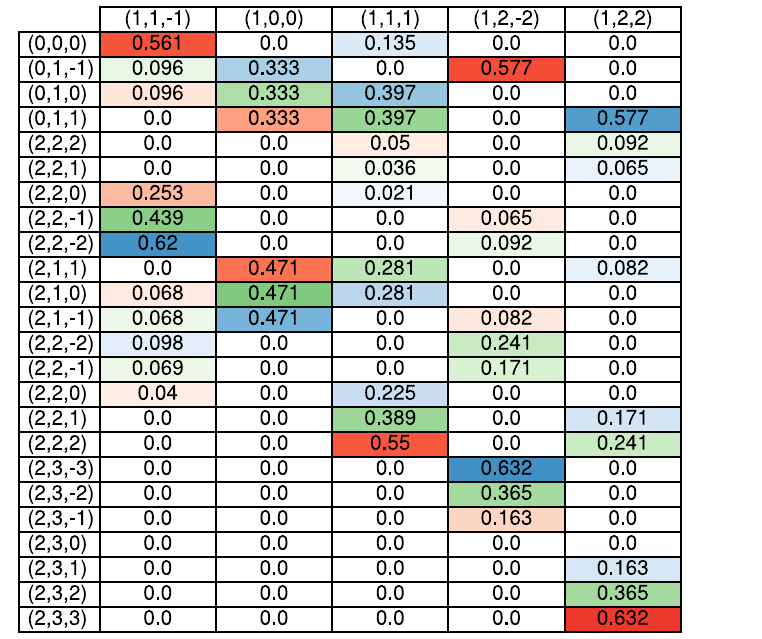}
    \centering
	\caption{Table showing the transition dipole moments in units of molecular dipole between some $N=1$ states, shown in columns and all $N=0,2$ states, shown in rows for CaF. Labels are in the form $(N, F, m_F)$. The relative intensity of each transition is given by the cell value, and highlighted by the colour, with darker colours indicating stronger transitions. Red, green and blue are chosen to represent coupling fields with polarisations that drive $\sigma^+, \pi$ and $\sigma^-$ transitions respectively.}
	\label{fig:tdm}
\end{figure}

\subsection{Transition Dipole Moments}
Rotational transitions in molecules can readily be driven by microwave fields. To understand the strength of these transitions one needs to calculate the transition dipole moment (TDM), which can be done using the function \texttt{transition\_dipole\_moment}. Figure \ref{fig:tdm} shows the TDM from a selection of $N=1$ states ($|1,1,-1\rangle, |1,0,0\rangle, |1,1,1\rangle, |1,2,-2\rangle$ and $|1,2,2\rangle$) to all states in rotational levels $N=0,2$ of CaF. Transitions from $N=1$ are considered as this is the state occupied following laser slowing. The strength of each transition is given in the cell, and highlighted by the colour. The red, green and blue sequential colour maps indicate the relative strengths for $\sigma^+, \pi$ and $\sigma^-$ transitions respectively. The calculations were done in the presence of an magnetic field of $60$~mG.

\section{Conclusion}
We have written a Python package to construct the Hamiltonian and calculate eigenenergies and eigenstates of $^2\Sigma$ state of molecules in the presence of external fields. The package is of particular relevance to molecules which can be laser cooled and we have provided constants for CaF, SrF and BaF molecules. As more molecular species are investigated, further constants can be added. Future iterations of the package could also be made to include an angle variable for magnetic and electric fields. 

\section{Code Availability and Installation}
The DiPolMol-Py files are available to download from the github repository \cite{DiPolMol}.

\section{Acknowledgements}
We thank Phil Gregory, Mike Tarbutt, Luke Caldwell, Jeremy Hutson and Simon Cornish for helpful discussions. Thank you also to Archie Baldock for testing of the package.\\
This work was supported by UKRI and EPSRC Grants: MR/X033430/1 and EP/X013758/1.

\appendix

\label{appA}
\section{}
Here we give the matrix elements for the various Hamiltonians included in the package. For brevity we use $|i\rangle$ as shorthand for the basis state $|N,J, F, m_{\rm F}\rangle$.

\subsection{Field-free Hamiltonian}
The rotational Hamiltonian is diagonal in our basis so can easily be written as
	
\begin{equation}
\begin{split}
		\langle i'|H_{\rm rot}|i\rangle = B_vN & (N+1)\delta_{N,N'}\delta_{J,J'}\delta_{F,F'}\delta_{m_{\rm F},m_{\rm F}'}\\ 
        & + D_v(N(N+1))^2\delta_{N,N'}\delta_{J,J'}\delta_{F,F'}\delta_{m_{\rm F},m_{\rm F}'}. 
\end{split}
\end{equation}

The hyperfine Hamiltonian can be divided into four terms which are expressed as:

\begin{equation}
\begin{split}
		\langle i'|\gamma\boldsymbol{S}\cdot \boldsymbol{N}|i\rangle = \frac{\gamma}{2}(J(J+1)-N(N+1)-\frac{3}{4})\delta_{N,N'}\delta_{J,J'}\delta_{F,F'}\delta_{m_{\rm F},m_{\rm F}'}, 
\end{split}
\end{equation}

\begin{equation}
\begin{split}
   	\langle i'|(b+c/3)\boldsymbol{I} \cdot \boldsymbol{S}|i\rangle =(-1) & ^{J+J'+F+N} \delta_{N,N'}\delta_{F,F'}\delta_{m_{\rm F},m_{\rm F}'}  \\
		& \times \frac{3}{2}((2J+1)(2J'+1))^{\frac{1}{2}}
		\begin{Bmatrix} J' & \frac{1}{2} & F\\ \frac{1}{2}&J&1 \end{Bmatrix}, 
\end{split}
\end{equation}

\begin{equation}
    \begin{split}
          \langle i'|\frac{\sqrt{6}c}{3} T^2(C)\cdot T^2(\boldsymbol{I}, \boldsymbol{S})|i\rangle = & \sqrt{10}c(-1)^{J+F+N'+\frac{1}{2}}\delta_{F,F'}\delta_{m_{\rm F},m_{\rm F}'}((2N+1)(2N'+1)\\
          &\times (2J+1)(2J'+1))^{\frac{1}{2}}
		\begin{Bmatrix} J'& \frac{1}{2} &F \\ \frac{1}{2} & J & 1 \end{Bmatrix}\
		\begin{Bmatrix} J'& J & 1 \\ \frac{1}{2} & \frac{3}{2} & N \end{Bmatrix}\\\
		&\times \begin{Bmatrix} N& N' & 2 \\ \frac{1}{2} & \frac{3}{2} & J' \end{Bmatrix}\
		\begin{pmatrix} N'& 2 & N \\ 0 & 0 & 0 \end{pmatrix},		
    \end{split}
\end{equation}

\begin{equation}
    \begin{split}
        \langle i'|c_{\rm F}\boldsymbol{I}\cdot \boldsymbol{N}|i\rangle =
        		(-1) & ^{2J'+F+N}\delta_{N,N'}\delta_{F,F'}\delta_{m_F,m_F'}\\
                & \times\left(\frac{3}{2}(2J+1)(2J'+1)N(N+1)(2N+1)\right)^{\frac{1}{2}}\\
        		& \times\begin{Bmatrix} J'& \frac{1}{2} &F \\ \frac{1}{2} & J & 1 \end{Bmatrix}\
        		\begin{Bmatrix} N'& J' & \frac{1}{2} \\ J & N & 1 \end{Bmatrix}.
    \end{split}
\end{equation}

\section{External field Hamiltonian}
Next we consider the Zeeman Hamiltonian which can decomposed into four terms in the following way for a magnetic field aligned along z. The second term is most naturally evaluated in Hund's case (a) basis, with the projections of $L$ and $S$ along the internuclear axis being given by $\Lambda$ and $\Sigma$ and we can define  $\Omega = \Lambda + \Sigma$. As $L=0$ we also have $\Lambda = 0$, this means that $\Omega = \Sigma = \pm\frac{1}{2}$. We then transform the basis back into Hund's case (b). The four terms are
\begin{equation}
    \begin{split}
        \langle i'|(g_s\mu_{\rm B} + g_l\mu_{\rm B})\boldsymbol{S}\cdot \boldsymbol{B}|i\rangle = & B_z\sqrt{\frac{3}{2}}(g_s\mu_{\rm B} + g_l\mu_{\rm B})\times (-1)^{(J'+J + F' + F-m_F+3+N')}\\
    & \times \delta_{N,N'} ((2F+1)(2F'+1)(2J+1)(2J'+1))^\frac{1}{2}\\
       & \times \begin{pmatrix}F' & 1 & F\\ -m_F' & 0 & m_F\end{pmatrix}
		\begin{Bmatrix}J' & F' & \frac{1}{2} \\ F & J & 1 \end{Bmatrix}\begin{Bmatrix}\frac{1}{2} & J' & N\\J &\frac{1}{2} & 1\end{Bmatrix},    
    \end{split}
\end{equation}
\begin{equation}
    \begin{split}
       	\langle i'|g_l\mu_{\rm B}(\boldsymbol{S}\cdot \lambda)
        (\boldsymbol{B}\cdot\lambda)|i\rangle =& B_z  g_l\mu_{\rm B} (-1)^{(F+F'-m_F'+2J'+N+N'+\frac{1}{2})}\\
        &\times\left((2F+1)(2F'+1)(2J+1)(2J'+1)\right)^\frac{1}{2}\\
		&\times\left((2N+1)(2N'+1)\right)^\frac{1}{2}
		\begin{Bmatrix}J & F & \frac{1}{2} \\F' & J' & 1 \end{Bmatrix}
        \begin{pmatrix}F' & 1 & F \\ -m_F' & 0 & m_F \end{pmatrix}\\
        &\times\sum_{\Omega = -\frac{1}{2}}^{\frac{1}{2}}(-1)^\Omega \begin{pmatrix}J' & 1 & J \\ -\Omega & 0 & \Omega \end{pmatrix} \begin{pmatrix}J & S & N \\ \Omega & -\Omega & 0 \end{pmatrix} \begin{pmatrix}J' & S' & N' \\ \Omega & -\Omega & 0 \end{pmatrix}\Omega, 
    \end{split}
\end{equation}
\begin{equation}
    \begin{split}
        \langle i'|(g_r\mu_{\rm B})\boldsymbol{N}\cdot \boldsymbol{B}|i\rangle =& B_z~g_r\mu_{\rm B}\times (-1)^{(J + J' + F + F'-m_F'+3 + N')}\delta_{N,N'} \\
        &\times\left((2F+1)(2F'+1)(2J+1)(2J'+1)\right)^\frac{1}{2}\\
        &\times\begin{pmatrix}F' & 1 & F\\ -m_F' & 0 & m_F\end{pmatrix}
		\begin{Bmatrix}J' & F' & \frac{1}{2} \\ F & J & 1 \end{Bmatrix}\begin{Bmatrix} N & J & \frac{1}{2} \\J' & N' & 1\end{Bmatrix}, 
    \end{split}
\end{equation}
\begin{equation}
    \begin{split}
        \langle i'|(g_N\mu_{\rm N})\boldsymbol{I}\cdot \boldsymbol{B}|i\rangle &= B_z~g_N\mu_{\rm N}\times (-1)^{(J' + 2F' -m_F'+3/2)}\delta_{J,J'}\\
        &\times\left(\frac{3}{2}(2F+1)(2F'+1)\right)^\frac{1}{2}\\
		&\times\begin{pmatrix}F' & 1 & F\\ -m_F' & 0 & m_F\end{pmatrix}
		\begin{Bmatrix}\frac{1}{2} & F & J \\ F' & \frac{1}{2} & 1 \end{Bmatrix}.
    \end{split}
\end{equation}

Under application of an external electric field of amplitude $E$ the dc electric field Hamiltonian is represented as
\begin{equation}
\begin{split}
	\langle i'| \mu_e \boldsymbol{E}\cdot \boldsymbol{z}|i\rangle =& \mu_e E (-1)^{(F'-m_F'+F+J'+ J+2N'+1)}((2F+1)(2F'+1)\\
    &\times(2J+1)(2J'+1)(2N+1)(2N'+1))^\frac{1}{2}\\
	&\times \begin{Bmatrix}J & F & \frac{1}{2} \\ F' & J' & 1 \end{Bmatrix} \begin{Bmatrix}N & J & \frac{1}{2} \\ J' & N' & 1 \end{Bmatrix} \begin{pmatrix}F' & 1 & F \\ -m_F' & 0 & m_F \end{pmatrix} \\
    &\times\begin{pmatrix} N' & 1 & N \\ 0&0&0 \end{pmatrix}.
\end{split}
\end{equation}

The matrix elements for the ac electric field effect under an applied off-resonant light field with frequency $\omega_{\rm L}$ are derived in~\cite{Caldwell2020a} and given here. $D_{M,0}^2$ is the Wigner D-matrix allowing rotation of the polarisation of the light which is important for the anisotropic light shift only. The total light shift experienced by the molecule is a sum over $k=0,1,2$, given by

\begin{equation}
\begin{split}
	\langle i' | H^k_{\rm{ac}} | i\rangle =& D_{M,0}^2(0,\beta,0) \times (-1)^{(F'-m_F'+F-J'+J+1)}((2F+1)(2F'+1)\\\
	&\times(2J+1)(2J'+1)(2N+1)(2N'+1))^\frac{1}{2}
	\begin{Bmatrix} J' & F' & \frac{1}{2} \\ F & J & K \end{Bmatrix} \\
	&\times\begin{pmatrix}
		F' & K & F \\ -m_F' & P & m_F
	\end{pmatrix}\begin{pmatrix}
		J & \frac{1}{2} & N \\ -\frac{1}{2} & \frac{1}{2} & 0
	\end{pmatrix}
	\begin{pmatrix}
		J' & \frac{1}{2} & N' \\ -\frac{1}{2} & \frac{1}{2} & 0
	\end{pmatrix}\\
    &\times
	\begin{pmatrix}
		J' & K & J \\ -\frac{1}{2} & 0 & \frac{1}{2} 
	\end{pmatrix} \alpha'_k.
\end{split}
\end{equation}

To calculate the components of the molecular polarisability we first calculate the molecule-frame parallel $\alpha_\parallel$ and perpendicular polarisability $\alpha_\perp$ components. In this package we consider only contributions from the X, A and B states. This gives us two expressions for $\alpha_\perp$ for $\Omega=\frac{1}{2}, \frac{3}{2}$, which are
\begin{equation}
    \alpha_\parallel = \sum_j\left(\frac{1}{\hbar(\omega_j+\omega_{\rm L})}+\frac{1}{\hbar(\omega_j-\omega_{\rm L})}\right) d_{X,j},
\end{equation}
\begin{equation}
    \alpha_{\perp,\Omega} = \sum_j\left(\frac{1}{\hbar(\omega_{k,\Omega}+\omega_{\rm L})}+\frac{1}{\hbar(\omega_j-\omega_{k,\Omega})}\right) {d}_{X,k},
\end{equation}
\begin{equation}
    \alpha_{\perp} = \frac{1}{2}\left(\alpha_{\perp, \frac{1}{2}}+\alpha_{\perp,\frac{3}{2}}\right),
\end{equation}
where $d_{X,j}$ and $d_{X,k}$ are the transition dipole moments between the $X(v=0)$ state and included excited states. These are summed over $j=X(v=1), B(v=0)$ and $k=A(v=0), A(v=1)$, respectively. The scalar, vector and tensor components of the polarisability can then be calculated. The equations describing these are
\begin{equation}
    \alpha_0 = \frac{1}{3}\left(\alpha_\parallel+2\alpha_\perp\right),
\end{equation}
\begin{equation}
    \alpha_1 = \frac{1}{2}\left(\frac{\omega_L}{\omega_\frac{1}{2}} \alpha_{\perp,\frac{1}{2}}+\frac{\omega_L}{\omega_\frac{3}{2}}\alpha_{\perp,\frac{3}{2}}\right),
\end{equation}
\begin{equation}
    \alpha_2 = \frac{2}{3}\left(\alpha_\parallel - \alpha_\perp\right).
\end{equation}

\bibliographystyle{elsarticle-num}
\bibliography{library2}

\end{document}